\documentclass[a4paper,12pt]{article}

\usepackage{amsmath, amssymb, amsfonts,amsthm}
\usepackage{graphicx, color, subfigure}
\usepackage{calc}
\usepackage{amsthm}
\usepackage{natbib}
\usepackage{mathdots}
\usepackage{enumitem}
\usepackage{standardmacros}
\usepackage{setspace, authblk}

\title{Spatial Backfitting of Roller Measurement Values from a Florida Test Bed}
\author[1]{Daniel K. Heersink\thanks{To whom correspondence should be addressed. Email: daniel.heersink@gmail.com}} \author[1]{Reinhard Furrer}
\author[2]{Mike A. Mooney}
\affil[1]{Institute of Mathematics, University of Zurich, CH-8057 Zurich}
\affil[2]{Colorado School of Mines, Golden, CO 80401, USA}
\date{}

\begin{document}
\maketitle
\doublespacing
\begin{quote}
  Modern earthwork compaction rollers collect location and compaction information as they traverse a compaction site. These data are indirectly observed through non-linear measurement operators, inherently multivariate with complex correlation structures, and collected in huge quantities. The nature of such data was investigated at a large, atypically compacted test bed in Florida, USA. Exploratory analysis of this data through detrending and empirical semivariogram estimation is performed. A second analysis using a sequential, spatial backfitting algorithm is used to investigate the importance of driving direction of the roller.
\end{quote}

\section{Modern Earthwork Compaction}
Modern compaction rollers monitor soil properties by observing stiffness characteristics of the soil. A vibrating drum traverses the compaction site at approximately 1m/s, compacting approximately 20cm of material at a time. Common construction practice is to compact several layers of material during the construction of a new road. Each layer is compacted in several passes of the roller until sufficient compaction is achieved. 

\begin{figure}[ht]
  \centering
  \includegraphics[width=9cm]{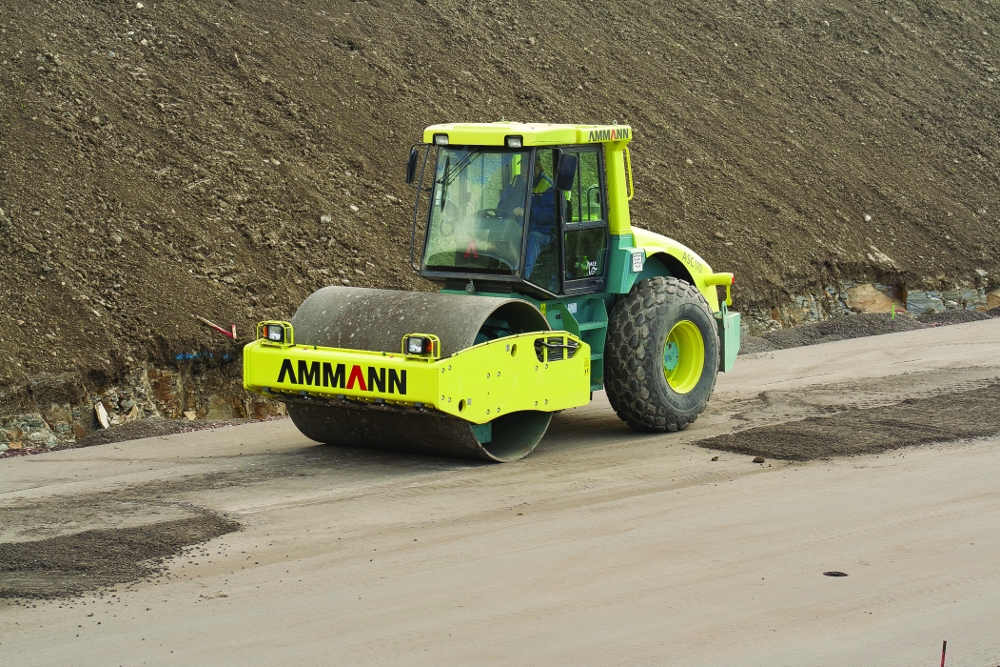}
  \caption{Ammann roller at work.} \label{fig:roller}
\end{figure}

Typical construction practice is to compact in segments of road 10--15m wide and 50--100m long. The roller traverses the compaction site in a snaking motion of several adjacent lanes. In practice, there is very little overlap between lanes \citep{Moon:etal:10}. See Figure \ref{fig:roller} for a typical compaction roller manufactured by Ammann.

\subsection{Roller Measurement Values (RMVs)}
A typical smooth drum has a diameter of approximately 1m and is approximately 2m long. An on-board sensor and GPS system record measurements that are together termed the roller measurement value (RMV). An individual RMV is an aggregate measure of a bulb of soil extending to a depth of approximately 1m with a diameter of 0.5--0.6m \citep{Faca:09}.

The physical nature of driving the roller down a lane with its vibrating drum causes other vibrational ``wobbling'' that remains fairly uniform over the course of the entire lane. Any bias this action produces will therefore be uniform over the entire lane. When the roller turns around and makes another pass down a different lane, the ``wobbling'' effect may be different though. This will lead to a change in the bias in the transverse direction, but the driving direction should remain unchanged as the new bias will be uniform over that entire lane. This is a cause of potential measurement error found only in the transverse direction.

\subsection{Florida Test Bed Data}
For a detailed investigation of roller properties, statistical characteristics, etc., a test bed with atypical dimensions was atypically densely compacted. A compaction roller traversed the compaction site in both the $x$- and $y$-directions. This Florida dataset consists of 19,145 observations of $x$- and $y$-coordinates, soil stiffness ($k_s$), and lane number in the $x$-direction driving and 19,975 observations in the $y$-direction driving. This analysis focuses on the driving direction.

\begin{figure}[ht]
  \centering
  \mbox{\subfigure{\includegraphics[width=7cm]{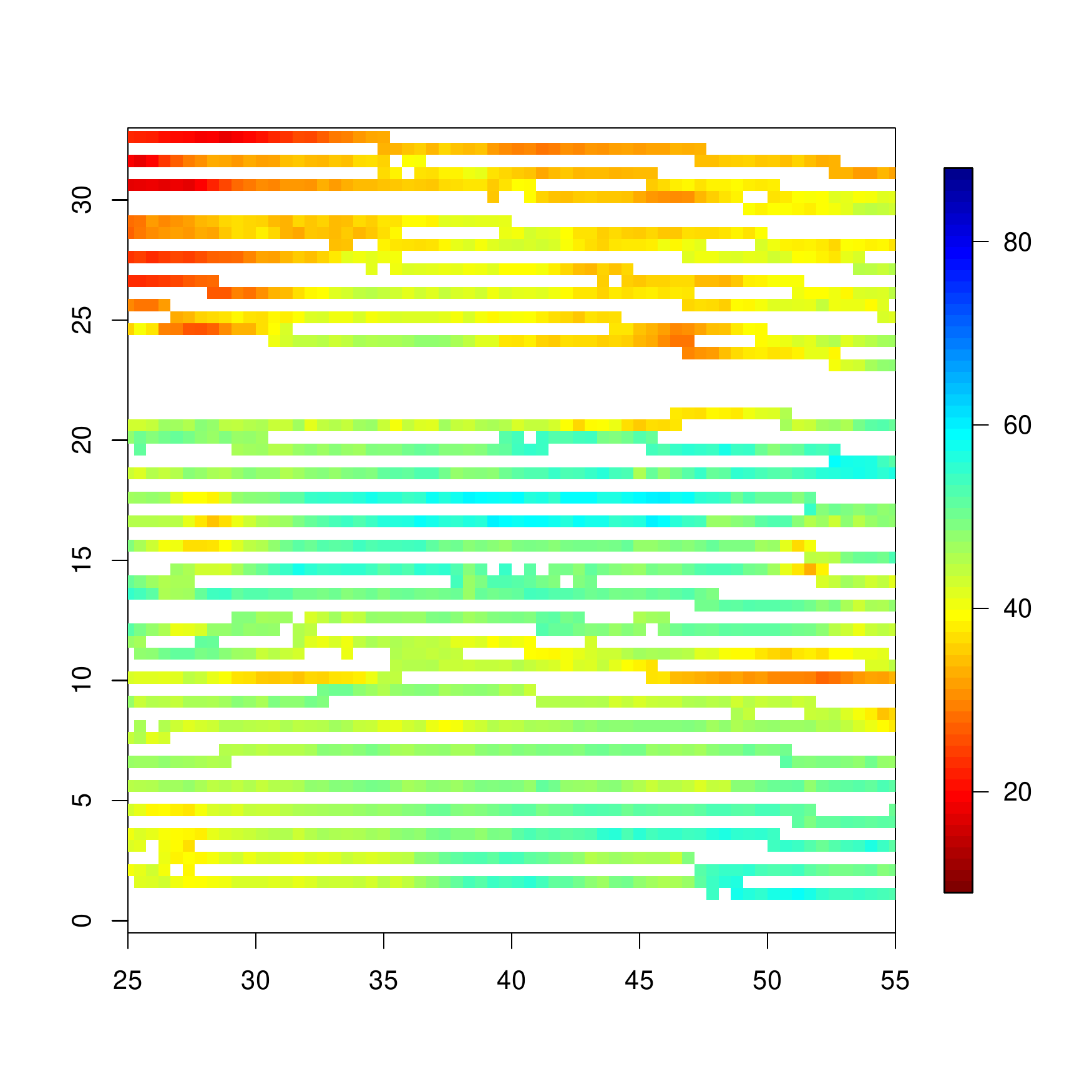}}\quad
  \subfigure{\includegraphics[width=7cm]{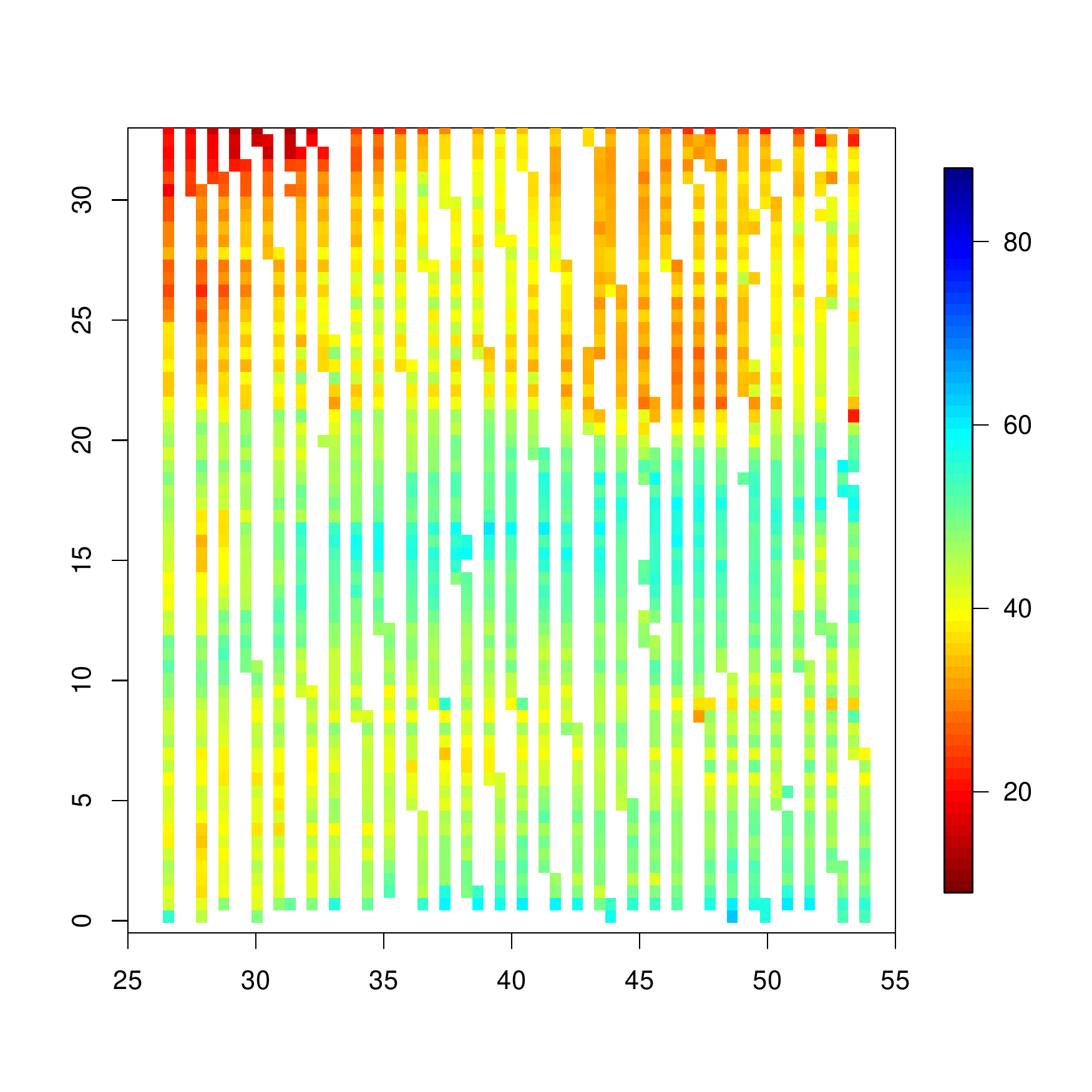} }}
  \caption{Data from the test bed in Florida, USA. RMVs collected from driving in the $x$-direction (left) and from driving in the $y$-direction (right) are depicted.} \label{fig:quilt}
\end{figure}

The roller first traversed the compaction site in the $x$-direction in a snaking fashion, first left-to-right and then back again right-to-left. The roller then traversed the compaction site a second time in the $y$-direction. The physical limitations of the site prohibited a snaking traversal in the $y$-direction, so the roller moved from bottom to top only. There are 29 lanes in the $x$-direction and 27 lanes in the $y$-direction. Figure \ref{fig:quilt} is a plot of the RMVs in the $x$-driving direction and the $y$-driving direction. Blue values represent high stiffness and red values represent low. An optimally compacted site would be uniformly blue.

\section{Exploratory Data Analysis}
\begin{figure}[ht]
  \centering
  \includegraphics[width=9cm]{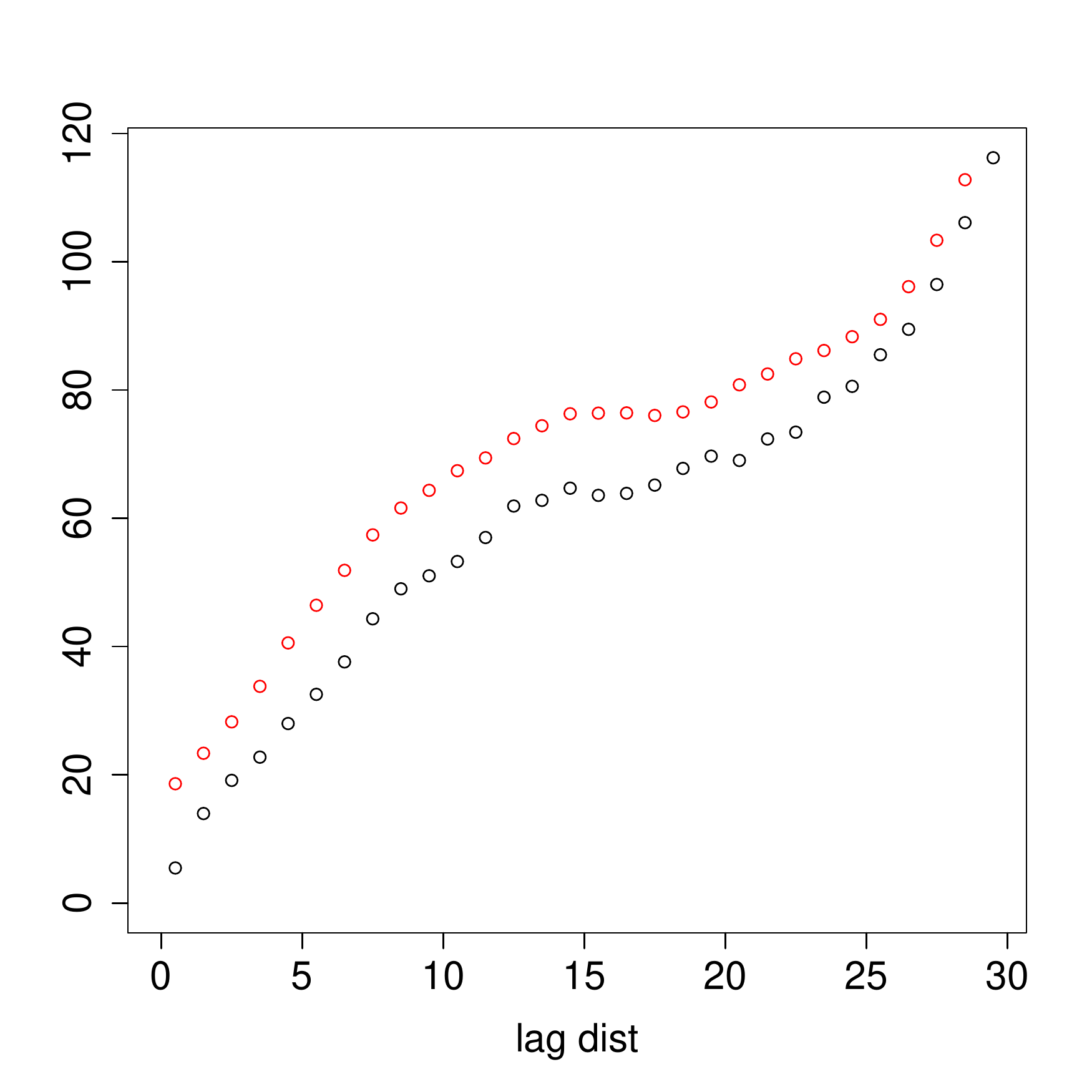}
  \caption{Empirical semivariograms of raw RMV data for $x$-direction driving (black) and $y$-direction driving (red).} \label{fig:firstvar}
\end{figure}

For this analysis, the $x$- and $y$-direction driving data are treated as two separate datasets. First, empirical semivariograms of the raw data were calculated using a subsample for computational reasons. These semivariograms exhibit aspects of non-stationarity. See Figure \ref{fig:firstvar} for representative empirical semivariograms of both driving directions.

\subsection{Detrending the Data}
The raw data exhibits a mean trend that must be removed as a constant mean is required to attain second-order stationarity. By detrending the data, we can remove the mean trend and proceed with the analysis utilizing a second-order stationary spatial process as a model.

\subsubsection{Small and Large Scale Variation}
Often times spatial data is modeled as 
\begin{equation}
  y(\s) = \mu(\s) + \alpha(\s) + \varepsilon(\s), \label{model}
\end{equation}
where $bmu(\s)$ is the mean structure of the process, $\alpha(\s)$ is the stochastic dependence structure of the process, and $\varepsilon(\s)$ is the measurement error. The mean structure is termed large scale variability and the dependence structure is termed small scale variability. What is termed mean structure and what is termed covariance structure is largely discretionary \citep{Cres:93}.

\subsubsection{Detrending Methods} \label{sec:mods}
Assuming model (\ref{model}), we desire a second-order stationary process $\alpha(\s)$. Therefore, the data detrending process should leave some structure in the data or all that will be left is the noise process $\varepsilon(\s)$, which is assumed uncorrelated. We have an assumption of spatial correlation.

The natural first choice for detrending is fitting a linear model: $\mu(\s) = \X\bbeta$. The residuals of the linear model fit can then be used to estimate the semivariogram of the stochastic structure terms $\alpha(\s) + \varepsilon(\s)$. The detrending process used included all cross products of the $x$- and $y$- coordinates. That is, for a 4th order polynomial, all products of $x$ and $y$ with a combined power of 4 or less were used. Empirical semivariograms were then calculated on the residuals of the linear model. Using a polynomial detrending of a 5th power generates empirical semivariograms with qualitatively identifiable nugget, partial sill and range parameters. This degree of detrending is desirable as all spatial variation is not lost and a constant mean of the residuals has been attained.

A practical, physical explanation of the linear model parameters is not of importance. The goal of detrending is establishing a constant mean of the residuals, and interpretability of the model parameters is insignificant.

An alternative to linear regression for data detrending is to detrend the data using a nonparametric function. For this analysis, the implementation of local polynomial smoothing known as locally weighted scatterplot smoothing (loess) was used \citep{Clev:79}. The loess smoothing approach is based on a moving window. A polynomial is fit to the data in a window using robust methods. The fitted value is then the predicted response at the middle of the window. The window is then slid over the range of the data, repeating the fitting process as the window moves \citep{Fara:06}. 

For this analysis, a span of 0.5 was used to reproduce empirical semivariograms like those of the polynomial detrend. This span corresponds to an estimated number of parameters of 13.5. This is approximately equivalent to a polynomial fit of 4th order, making this method comparable to that of a polynomial detrending.

\subsection{Fitting to a Model}
The empirical semivariograms calculated from the loess detrended data were then fitted to a spherical model with Cressie weights using the \texttt{variofit} function in \texttt{R}. The spherical model was chosen as the empirical semivariograms seemed to exhibit a linear behavior near the origin. The spherical model also induces sparse matrix structures, helpful for computation. The spherical model is defined as
\begin{align}
  C(h;\btheta) = \left\{ \begin{array}{rcl} \theta_0(1 - 1.5(h/\theta_1) + .5(h/\theta_1)^3) & \mbox{for} & h \in [0, \theta_1) \\ 0 & \mbox{for} & h \ge \theta_1 \end{array}\right. , \label{eq:sph}
\end{align}
where $\theta_0$ is the (partial) sill and $\theta_1$ is the range of the spatial process. 

Cressie weights were chosen because they are the most commonly used weights for fitting empirical semivariograms to a covariance model. Weighted least squares and generalized least squares require knowing the covariance structure of the semivariogram. While this is possible, it is hard to implement. \citet{Cres:85} proposed a weighting structure that is a compromise of weighted least squares that is no more difficult to compute than ordinary least squares.

\subsection{Semivariogram Uncertainty}
\begin{figure}[ht]
  \centering
  \includegraphics[width=9cm]{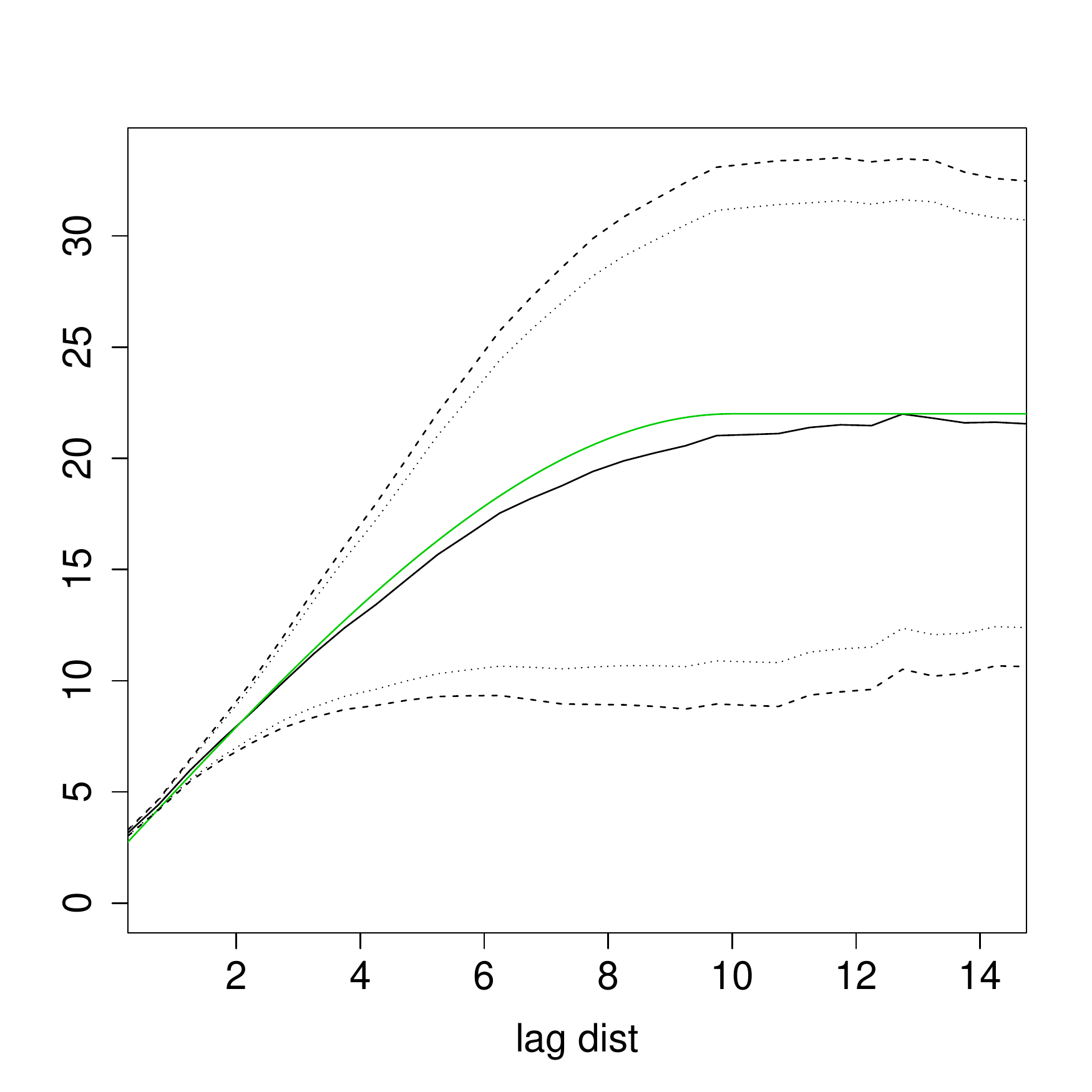}
  \caption{Mean of simulated empirical semivariograms (solid line) and 95\% (dashed line) and 75\% confidence bands (dotted line). The true spherical semivariogram is depicted in green.} \label{fig:uncert}
\end{figure}

Estimates of total sill, range, and nugget have very large confidence intervals. As the lag distance increases, the confidence interval for the total sill also increases \citep{Nord:Cara:08}. A simple simulation of several random fields with semivariogram parameters chosen to match those of the empirical semivariograms from this study was performed. From these random fields, empirical semivariograms were then calculated. A mean and standard deviation of these semivariograms was then calculated and these were used to calculate pointwise confidence intervals. The estimated confidence bound of the semivariogram starts very small for a lag distance of zero and begins expanding for larger lag distances. This expansion continues for larger lag distances. Decreasing the confidence to 75\% does very little to improve the width of the estimated confidence bounds for large lag distances, see Figure \ref{fig:uncert}.

\subsection{Sampling Concerns} \label{sec:samp}
To maintain computational efficiency, the data was subsampled for empirical semivariogram estimation. 10,000 data points were sampled from each of $x$- and $y$-direction driving datasets. A loess detrending of each sample was performed. This produced two detrended datasets from which subsamples of 2500, 3500, and 4500 data points were drawn. Directional empirical semivariograms were then calculated in the $x$- and $y$-direction to generate a total of twelve empirical semivariograms. These empirical semivariograms were then fit to a spherical model.

\begin{figure}
  \centering
  \mbox{\subfigure{\includegraphics[width=8cm]{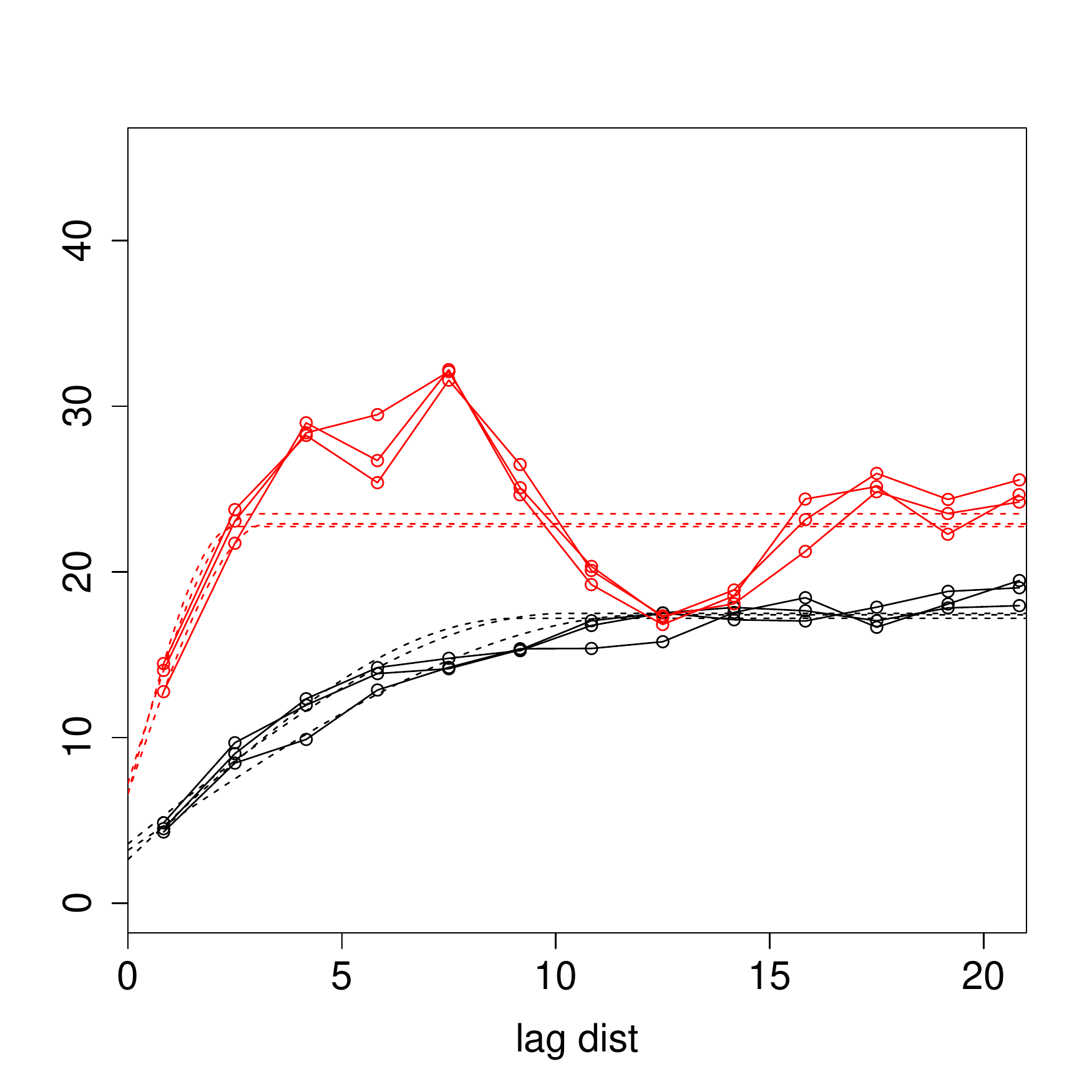}}\quad
  \subfigure{\includegraphics[width=8cm]{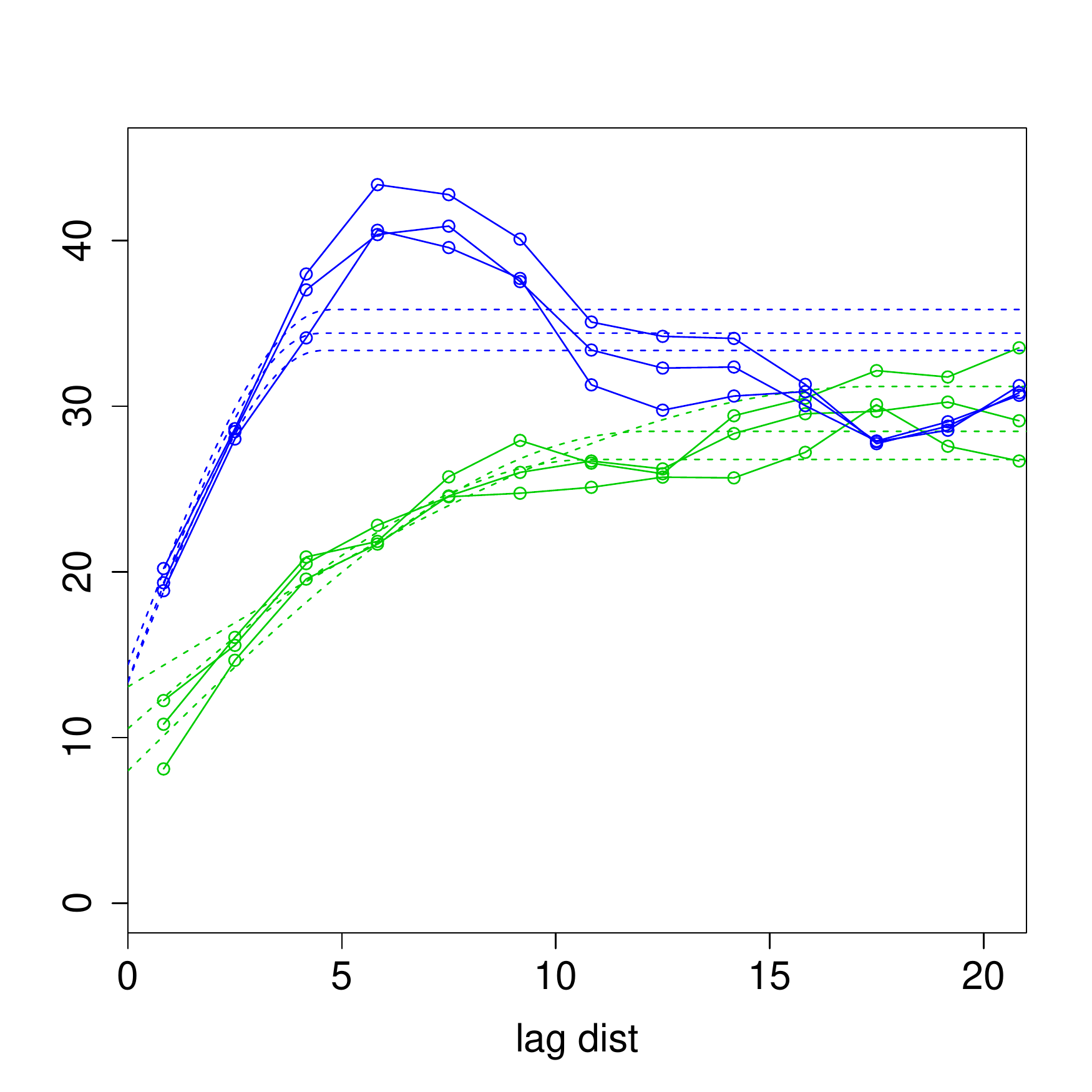} }}
  \caption{Directional empirical semivariograms and fitted spherical models from three polynomial detrended subsamples of RMVs of the $x$-driving direction subset (left) and of the $y$-driving direction subset (right). Dashed lines indicated fitted models, $x$-directional semivariograms are in black and green and $y$-directional semivariograms are in red and blue.}
  \label{fig:empvario}
\end{figure}

There was no discernible difference between the empirical semivariograms within each dataset. Figure \ref{fig:empvario} depicts the empirical and fitted directional semivariograms of the $x$-driving direction dataset (left) and $y$-driving direction (right). Since the sampled directional empirical semivariograms are essentially identical within each dataset, we concluded the subsampling was adequate, i.e. the subsampling produced a representative sample

\subsection{Results}
The next step is a qualitative analysis of the characteristic semivariogram features. Semivariograms for both driving directions exhibit similar features. For both driving directions, the $y$-directional semivariograms have a range of 0--5 and the $x$-directional semivariograms exhibit a range of 9--15. 

For the $x$-driving direction, the total sill for $y$-directional semivariograms is 22--25, and 15--18 for $x$-directional. In the $y$-driving direction, the total sill is 33--36 for $y$-directional and 26--30 for $x$-directional semivariograms. Similarly, for $x$-driving direction, the nugget for $y$-directional semivariograms is 5--8 and 2--5 for $x$-directional. For the $y$-driving direction, the nugget for $y$-directional semivariograms is 10--15 and 10--12 for $x$-directional semivariograms, see Table ~\ref{tab:contr} and Figure \ref{fig:empvario}.

\begin{table}[ht]
  \caption{Directional semivariogram parameters \label{tab:contr}}
\medskip
  \centering
  \begin{tabular}{l|cc|cc|}
    & \multicolumn{2}{|c|}{$x$-driving} & \multicolumn{2}{|c|}{$y$-driving}\\\hline
    & $x$-directional & $y$-directional & $x$-directional & $y$-directional\\\hline
    range & 9--15 & 0--5 & 9--15 & 0--5 \\
    total sill & 15--18 & 22--25 & 26--30 & 33--36 \\
    nugget & 2--5 & 5--8 & 10--12 & 10--15 \\\hline
  \end{tabular}
\end{table}

\section{Anisotropy Concerns} \label{sec:anisotropy}
Based on these observations, it is fairly safe to assume that there is no sill or nugget anisotropy. There does appear to be a range anisotropy between the $x$-directional semivariograms and the $y$-directional semivariograms. The ratio of the range in the $x$-direction vs. the $y$-direction is approximately 5:1. The empirical semivariograms indicate a geometric range anisotropy that can be dealt with by a simple transformation of the data locations, \citep{Zimm:93}. 

This geometric range anisotropy can possibly be explained by the compaction process. As the roller traverses the compaction site, it collects data every 10cm in the driving direction. Data is collection in the direction perpendicular to the driving direction approximately every 1--2m. The vibrating drum is approximately 2m wide, thus the $y$-directional location of observations in adjacent lanes is 1--2m apart, dependent on the placement of the GPS unit. Also, material is brought into the compaction site via a dump truck and laid down in sections. It is unclear if discontinuities exist on the boundaries of these sections. If they do exist, they could contribute to range anisotropy.

Due to the nature of the driving process, data points are much more closely spaced in the driving direction than they are in the transverse direction. This leads to difficulties estimating the nugget in the transverse direction as the smallest lag distance is on the order of 1--2m. The nugget anisotropy could therefore be explained by a vertical shift of the entire semivariogram caused by a measurement error in the transverse direction. This would essentially be a nugget anisotropy model.

Let the true compaction process be denoted by $Z(x,y)$ and the data we collect be denoted by $Y(x,y) = Z(x,y) + \varepsilon(y)$, where $\varepsilon(y)$ is a measurement error seen only in the $y$-direction. Then, the semivariogram of the $Y$ process is $\gamma_Y(h_x,h_y) = \var(Y(x,y) - Y(x+h_x,y+h_y)) = \var(Z(x,y) - Z(x+h_x,y+h_y) + \varepsilon(y) - \varepsilon(y+h_y)) = \gamma_Z(h_x,h_y) + \gamma_{\varepsilon}(h_y)$. The $x$-directional semivariogram is then $\gamma_x(h_x) = \gamma_Y(h_x,0) = \gamma_Z(h_x,0)$ and the $y$-directional semivariogram is $\gamma_y(h_y) = \gamma_Z(0,h_y) + \gamma_{\varepsilon}(h_y)$. Thus the transverse directional semivariogram is shifted up by the measurement error $\varepsilon$.

\section{Driving Direction Investigation}
We utilize a state-space formulation to handle unique observation locations. Assume the RMVs can be decomposed into an underlying mean trend dependent on spatial location, driving direction, speed, and vibration amplitude, and a Gaussian spatial random process, (i.e. $\w = \X\bbeta + \balpha$), where the domain of $\w$ is a lattice. Here, $\X$ is a full rank matrix of the fixed effects covariates and $\balpha$ represents an unknown, spatially varying random process. The observed locations of the RMVs are then mapped to the lattice. 

Implementing a sequential, spatial mixed-effects model \cite{Heer:Furr:13}, we can model the Florida dataset as:
\begin{align*}
  \z_x &= \H_x\X_x\bbeta_x + \H_x\balpha_x + \bvarepsilon_x \\
  \z_y &= \H_y\X_y\bbeta_y + c\H_y\balpha_x + \H_y\balpha_y + \bvarepsilon_y,
\end{align*}
where $\balpha_x$ and $\balpha_y$ correspond to random variation of the layer of material being compacted during driving in the $x$- and $y$-direction, $\H_x$ and $\H_y$ are the operators mapping lattice points to observed locations, and $\bvarepsilon_x$ and $\bvarepsilon_y$ represent the measurement error of the sensor. For this analysis, the lattice chosen is an $80 \times 80$ grid of points equally spaced on $[25, 55] \times [-.5, 33]$. The size of the grid was chosen to encompass all observation locations. 

We also utilize a range anisotropy parameter $\rho$, given the empirical semivariograms calculated in Section \ref{sec:samp}. The range anisotropy is handled with a transformation of the coordinates. Thus $\rho$ is the ratio of the range in the $x$-direction to that in the $y$-direction and the transformation matrix $\A$ is defined as $\A = \diag(1, \rho)$.

As detailed in \cite{Heer:Furr:13}, any additive term that can be estimated in a mathematically equivalent way as universal kriging can also be included in such a model. Splines are such an additive component that has this mathematical equivalency. The literature on splines is extensive and computational feasibility can be maintained, i.e. \cite{Wahb:90}, \cite{Eile:etal:96}, \cite{Marx:Eile:98}, \cite{Eile:Marx:04}.

Since there was not a new layer of material added to the compaction site after compacting in the $x$-direction, the measurements in the $y$-direction are measurements of the same process as those in the $x$-direction. Thus, we should expect $c \rightarrow 1$ and either $\balpha_y \rightarrow \0$ or $\balpha_y \rightarrow \bgamma_y$, where $\bgamma_y$ represents a spatially varying process only in the $y$-direction, e.g. a process representing the nugget anisotropy discussed in Section \ref{sec:anisotropy}. Thus we would expect to see an empirical semivariogram of $\balpha_y$ to either have a sill of zero or a very small range. Due to measurement errors, a pure nugget model is not expected.

The Sequential Backfitting Algorithm from \cite{Heer:Furr:13} was applied to the data, setting $c = 1$, with $p=2$, $\X$ is the fixed effects matrix containing all 5th degree and lower polynomial combinations of the centered and scaled $x$- and $y$-direction coordinates of the roller. To create sparse matrix structures and aid in computation a spherical covariance function was assumed, see equation (\ref{eq:sph}).

The semivariogram estimation done in this study is directional. The empirical semivariograms were calculated in the driving direction. Thus, for $\balpha_x$ empirical semivariograms were calculated in the $x$-direction and in the $y$-direction for $\balpha_y$.

\section{Backfitting Results}
\begin{figure}
  \centering
  \mbox{\subfigure{\includegraphics[width=8cm]{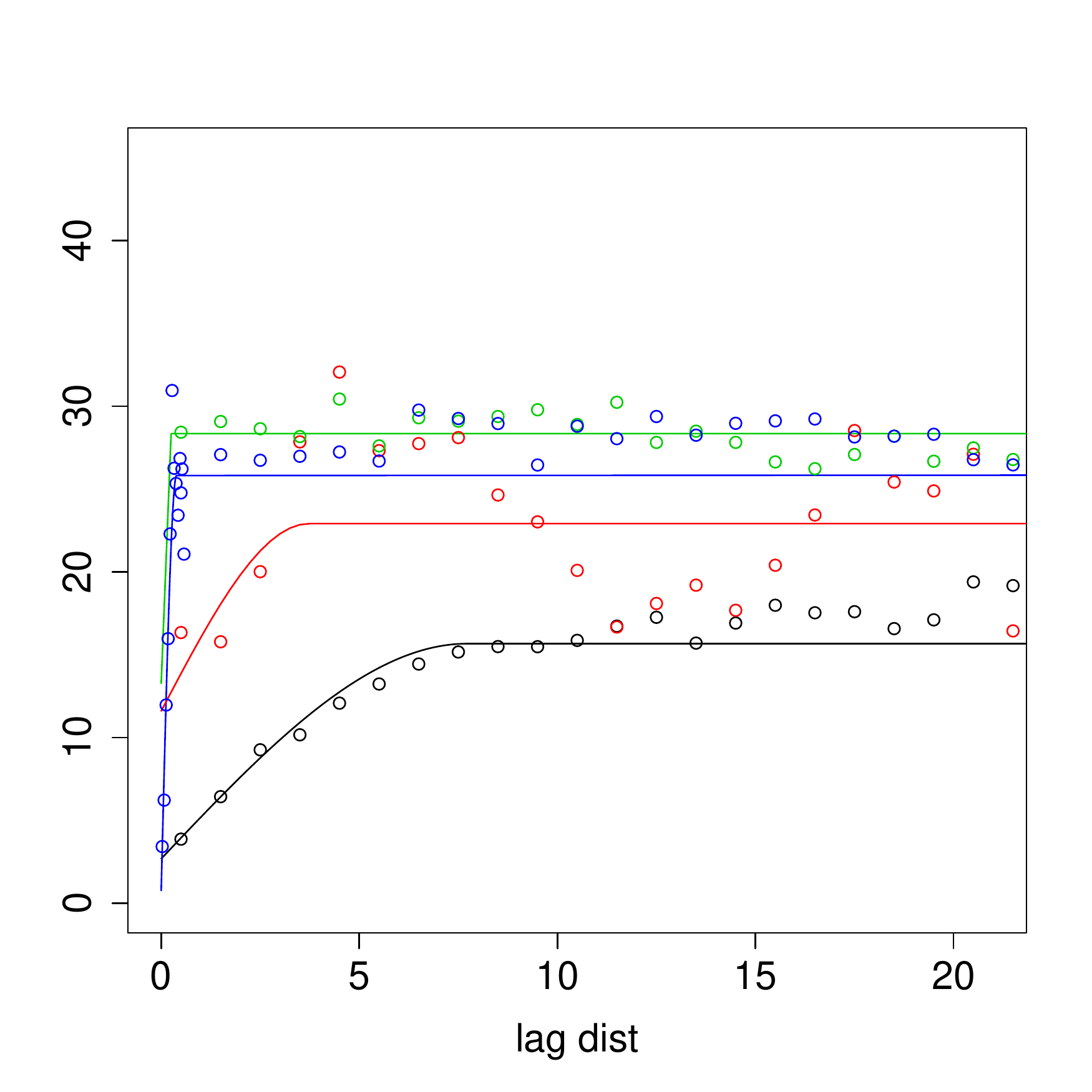}}\quad
  \subfigure{\includegraphics[width=8cm]{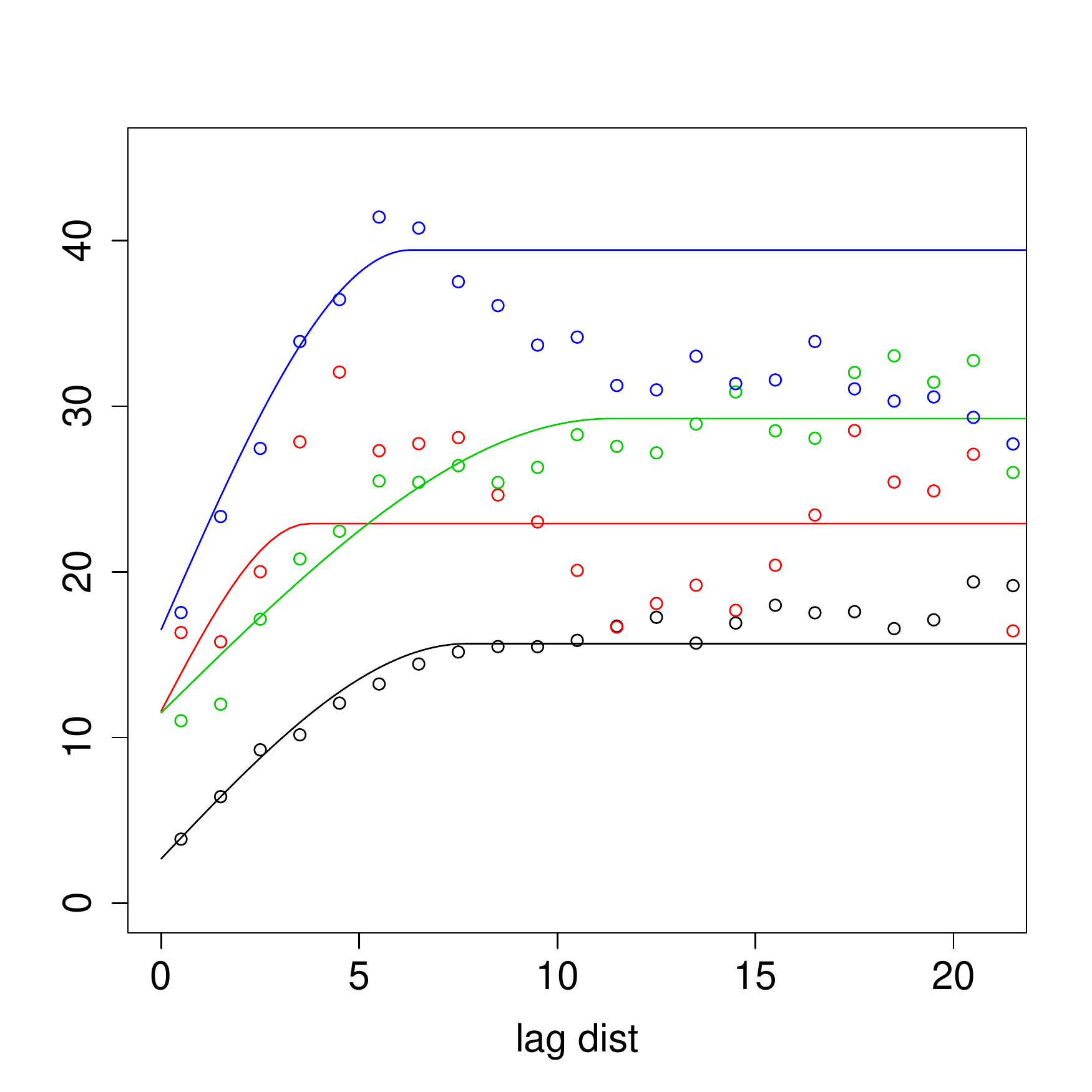} }}
  \caption{Fitted $x$-directional semivariograms for $\balpha_x + \bvarepsilon_x$ (black) and $\balpha_y + \bvarepsilon_y$ (green) and $y$-directional semivariograms for for $\balpha_x + \bvarepsilon_x$ (red) and $\balpha_y + \bvarepsilon_y$ (blue) for $c=1$ (left) and $c=0$ (right). The $c=0$ plot reproduces the curves from Figure \ref{fig:empvario}, as would be expected.}
  \label{fig:back}
\end{figure}

The estimated covariance parameters for $\balpha_x$ are $\widehat\btheta_x = (12.99, 7.72)\T$ and $\widehat\btheta_y = (25.08, 0.39)\T$ for $\balpha_y$, the estimated variances of $\bvarepsilon_x$ and $\bvarepsilon_y$ are $\widehat{\sigma}_x^2 = 2.68$ and $\widehat{\sigma}_y^2 = 0.75$, see Figure \ref{fig:back}. The backfitting procedure thus reproduces the empirical $x$-directional semivariogram from the standard detrending approach. 

The range of the $\balpha_y$ process is relatively small, thus there is no evidence to reject the assumption that $\balpha_y \rightarrow \bgamma_y$ from this analysis. This backfitting analysis thus reconfirms the existence of a nugget effect in the $y$-direction. This would imply the ``static'' rolling done by the roller after compaction was completed is generally truly static and the material is not being actively compacted during this phase of construction. 

The backfitting procedure was also run for $c=0$. As can be seen in the right plot of Figure \ref{fig:back}, the calculated semivariograms are reproductions of the standard detrending approach of Section \ref{sec:mods} and the empirical semivariograms found in Figure \ref{fig:empvario}.

\bibliographystyle{mywiley}
\bibliography{mybib}

\end{document}